\documentclass[prc,aps,floatfix,showpacs,twocolumn,nofootinbib]{revtex4-1}
\usepackage{dcolumn}
\usepackage{slashed}
\usepackage{bm}
\usepackage{color}
\usepackage{graphicx}
\usepackage{amssymb,amsmath}
\usepackage[utf8]{inputenc} 							
\usepackage{bbm}
\usepackage{simplewick}
\usepackage{float}
\usepackage{simplewick}
\usepackage{xcolor}
\usepackage[normalem]{ulem}

\def\bmp{{\bm p}}
\def\bmk{{\bm k}}

\DeclareMathOperator{\nrd}{\overleftrightarrow{\nabla}}

\begin{document}
\title{Effect of the N3LO three-nucleon contact interaction on $p-d$ scattering observables}

\author{ L. Girlanda$^{1,2}$, E. Filandri$^{1,2}$, A. Kievsky$^3$,
L.E. Marcucci$^{3,4}$ and  M. Viviani$^3$}

\affiliation{
$^1$Department of Mathematics and Physics ``E. De Giorgi", University of Salento, I-73100 Lecce, Italy \\
$^2$INFN-Lecce, I-73100 Lecce, Italy \\  
$^3$INFN-Pisa, I-56127, Pisa, Italy \\
$^4$Department of Physics ``E. Fermi'', University of Pisa, I-56127 Pisa, Italy}
\date{\today}

\begin{abstract}
A unitary transformation allows to remove redundant terms in the two-nucleon (2N) contact interaction at the fourth order (N3LO) in the low-energy expansion of Chiral Effective Field Theory. In so doing a three-nucleon (3N) interaction is generated. We express its short-range component in terms of five combinations of low-energy constants (LECs) parametrizing the N3LO 2N contact Lagrangian. Within a hybrid approach, in which this interaction is considered in conjunction with the phenomenological AV18 2N potential, we show that the involved LECs can be used to fit very accurate data on polarization observables of low-energy $p-d$ scattering, in particular the $A_y$ asymmetry. The resulting interaction is of the right order of magnitude for a N3LO contribution.

\end{abstract}

\pacs{}

\maketitle
\section{Introduction}
The effective field theory (EFT) framework is nowadays the standard setting to address the nuclear interaction problem. Starting from the choice of low-energy active degrees of freedom, it allows to express physical observables in terms of low-energy constants (LECs) in a systematic expansion in powers of a small parameter representing the separation of scales \cite{Bedaque_2002,Epelbaum_2006,Epelbaum_2009,Machleidt_2011}.
All the short-distance effects from the frozen degrees of freedom are effectively encoded in the values of the LECs, which parametrize contact interactions. Being unconstrained by the imposed symmetries, the LECs have to be fitted from experimental data, at the cost of the predictive power of the EFT. From another perspective, they may provide the needed flexibility to accurately model the nuclear interaction. 
For example, Chiral EFT (ChEFT) makes use of the approximate chiral symmetry of strong interaction to severely constrain the interaction of pions and nucleons. According to the common wisdom, in the isospin limit there are two LECs contributing at the leading order (LO) in the two-nucleon (2N) sector, traditionally called $C_S$ and $C_T$, seven ($C_{i=1,...,7}$) at the next-to-leading order (NLO), and fifteen more ($D_{i=1,...,15}$) at the next-to-next-to-next-to leading order (N3LO). The three-nucleon (3N) sector is much more constrained. The first contributions arise at the next-to-next-to leading order (N2LO) \cite{Epelbaum_2002,Navratil_2007}, parametrized by two LECs, $C_D$ and $C_E$, the former  of which is actually a weak two-nucleon LEC, contributing for example to the muon capture from deuterium \cite{gardestig,gazit,marcucci12,ceccarelli}.  At the fifth order (N4LO) we find the contribution of thirteen more LECs, $E_{i=1,...,13}$ \cite{Girlanda_2011}. However, the distinction between 2N and 3N  LECs is, to some extent, a matter of convention, depending on the arbitrary choice for the nucleon interpolating field: non-linear field redefinitions may change a seemingly  2N interaction into a 3N one, without changing the predictions for the on-shell quantities.
Thus, it was observed in Ref.~\cite{Reinert} that three out of the fifteen 2N independent contact interactions arising at N3LO can be made to vanish by a suitable unitary transformation, inducing specific modifications of the 3N interaction. In Ref.~\cite{Girlanda_2020} we exhibited the precise form of the induced 3N interaction. Moreover, we identified two additional 2N contact LECs at N3LO parametrizing momentum dependent interactions allowed by Poincar\'e symmetry, which we named $D_{16}$ and $D_{17}$. They can also be transformed, through a unitary transformation, into a 3N interaction. Thus, contrary to widespread belief \cite{PhysRevC.77.064004,Bernard_2011}, five adjustable LECs parametrize the 3N interaction at N3LO of the chiral expansion. This could be the explanation of all failed attempts to improve the accuracy in 3N systems (particularly for scattering observables) when going from N2LO to N3LO \cite{Golak}. On the other hand, the inclusion of the N4LO 3N contact interaction has already proved to be of great importance in reducing existing discrepancies between theory and experimental data \cite{contact19,witala}.
In the present paper we provide quantitative evidence that the five extra LECs at N3LO ensure sufficient flexibility to drastically improve the description of low-energy $p-d$ scattering polarization observables, most notably the $A_y$ asymmetry, which constitutes a long-standing problem for most nuclear interaction models. We do this in a hybrid approach in which the induced 3N force (restricted to its shortest range part) is considered in conjunction with the phenomenological AV18 2N potential \cite{av18}, deferring a consistent calculation in ChEFT to future work. 
The paper is organized as follows. In Section \ref{sec:2N} we start from the N3LO 2N Hamiltonian comprising 17 independent LECs and identify the unitary transformation allowing to restrict oneself to a subset of twelve independent LECs. In Section \ref{sec:3N} the corresponding induced 3N contact interaction are worked out and written in the minimal basis of thirteen independent subleading operators of Ref.~\cite{Girlanda_2011}. Despite being formally of N4LO, these specific terms are promoted in the chiral counting to N3LO, due to the enhancement produced by the dependence on the large nucleon mass. In Section \ref{sec:methode} we briefly describe the calculation of low-energy $p-d$ scattering observables below the breakup threshold based on the Hyperspherical Harmonics (HH) method \cite{hh1,hh2}. In Section \ref{sec:results} we fit the 5 N3LO LECs to  very precise data on polarized $p-d$ scattering at 2~MeV center-of-mass energy \cite{shimizu} using the induced 3N interaction on the top of the AV18 2N phenomenological potential. Finally, we summarize our findings and outline directions for future developments in the conclusions of Section~\ref{sec:conclusions}.
\section{ N3LO 2N contact Hamiltonian }\label{sec:2N}
The N3LO 2N contact potential was originally considered in Refs.~\cite{machleidt_plb02,machleidt_prc03} as consisting of 15 LECs. After careful scrutiny of the constraints imposed by Poincar\'e symmetry, two further LECs emerge \cite{Girlanda_2020}, leading to the following expression in the general reference frame,  
 \begin{align}
 V_{NN}^{(4)}=& D_{1} k^{4}+D_{2} Q^{4}+D_{3} k^{2} Q^{2}+D_{4}(\bm{k} \times \bm{Q})^{2} +\left[D_{5} k^{4} \right.\nonumber \\&\left. +D_{6} Q^{4}+D_{7} k^{2} Q^{2}+D_{8}(\bm{k} \times \bm{Q})^{2}\right]\left(\bm{\sigma}_{1} \cdot \bm{\sigma}_{2}\right) \nonumber\\
&+\frac{i}{2}\left(D_{9} k^{2}+D_{10} Q^{2}\right)\left(\bm{\sigma}_{1}+\bm{\sigma}_{2}\right) \cdot(\bm{Q} \times \bm{k})\nonumber\\&+\left(D_{11} k^{2}+D_{12} Q^{2}\right)\left(\bm{\sigma}_{1} \cdot \bm{k}\right)\left(\bm{\sigma}_{2} \cdot \bm{k}\right)\nonumber\\&+\left(D_{13} k^{2}+D_{14} Q^{2}\right)\left(\bm{\sigma}_{1} \cdot \bm{Q}\right)\left(\bm{\sigma}_{2} \cdot \bm{Q}\right) \nonumber\\&
+D_{15} \,\bm{\sigma}_{1} \cdot(\bm{k} \times \bm{Q}) \,\bm{\sigma}_{2} \cdot(\bm{k} \times \bm{Q})\nonumber\\&+i D_{16}\, \bm{k} \cdot \bm{Q}\, \bm{Q} \times \bm{P} \cdot\left(\bm{\sigma_{1}}-\bm{\sigma_{2}}\right)\nonumber\\&+D_{17}\, \bm{k} \cdot \bm{Q}\,(\bm{k} \times \bm{P}) \cdot\left(\bm{\sigma_{1}} \times \bm{\sigma_{2}}\right)\, \label{eq:2N}
\end{align}
with $\bm{ k}=\bm{ p}'-\bm{ p}$ and $\bm{ Q}=\frac{\bm{ p}'+\bm{ p}}{2}$, $\bm{ p}$ and $\bm{ p}'$ being the initial and final relative momenta, and $\bm{ P}=\bm{p}_1+\bm{ p}_2$ the total pair momentum.
However, as it was pointed out in Ref.~\cite{Reinert},  only 12 independent LECs survive on shell and can thus be determined from 2N scattering data.
This redundancy amounts to a unitary ambiguity, i.e. to the possibility of generating shifts of the LECs by unitary transforming the one-body kinetic energy operator $H_0$ as
\begin{equation}
    H_0 \to  U^\dagger H_0  U,
\end{equation}
where $U$ is the most general unitary 2-body contact transformation depending on 5 arbitrary parameters $\alpha_i$,
\begin{equation} \label{eq:unitaryh0}
    U = \exp \left[\sum_{i=1}^5 \alpha_i T_i\right],
\end{equation}
and the independent generators $T_i$ were given explicitly in Ref.~\cite{Girlanda_2020} as
\begin{eqnarray}
  T_1 &=& \int d^3{\bf x}\, N^\dagger \nrd^i N \nabla^i (N^\dagger N)  , \\
  T_2 &=& \int d^3{\bf x}\, N^\dagger \nrd^i  \sigma^j N \nabla^i  (N^\dagger \sigma^j N)  , \\
  T_3 &=& \int d^3{\bf x} \left[ N^\dagger \nrd^i  \sigma^i N \nabla^j  (N^\dagger \sigma^j N) \right. \nonumber \\
  && \left. + N^\dagger \nrd^i  \sigma^j N \nabla^j  (N^\dagger \sigma^i N) \right], \\
 T_4 &=&  i \epsilon^{ijk} \int d^3{\bf x} \,N^\dagger \nrd^i N  N^\dagger \nrd^j \sigma^k N  , \\   
 T_5 &=& \int d^3{\bf x} \left[ N^\dagger \nrd^i  \sigma^i N \nabla^j ( N^\dagger \sigma^j N) \right. \nonumber \\
 && \left.- N^\dagger \nrd^i  \sigma^j N \nabla^j ( N^\dagger \sigma^i N )\right],
\end{eqnarray}
with $N^\dagger \nrd^i N = N^\dagger (\nabla^i N) -(\nabla^i N^\dagger ) N$, and $N(x)$ denoting the non-relativistic nucleon field operators.

The transformation (\ref{eq:unitaryh0}) entails a shift of the N3LO contact LECs, $D_i \to D_i + \delta D_i$, with
\begin{align}
\delta D_3 &= - \frac{4}{m}\alpha_1, \\
\delta D_4 &= \frac{4}{m}\alpha_1, \\
\delta D_7 &=  -\frac{4}{m}\alpha_2, \\
\delta D_8 &=\frac{4}{m}\alpha_2 +\frac{2}{m}\alpha_3, \\
\delta D_{15} &= - \frac{4}{m}\alpha_3,\\
\delta D_{12}& = - \frac{4}{m}\alpha_3, \\
\delta D_{13} &= -\frac{4}{m}\alpha_3, \\
\delta D_{16} &= -\frac{2}{m}\alpha_4  ,\\
 \delta D_{17} &= -\frac{4}{m}\alpha_3 - \frac{2}{m}\alpha_5,
\end{align}
and the remaining ones being zero. Here $m$ is the nucleon mass.
By choosing  
\begin{align} 
\alpha_1 &= \frac{m}{16} \left( 16 D_1 + D_2 + 4 D_3\right), \label{eq:alpha1}\\
\alpha_2 &=\frac{m}{16} \left( 16 D_5 + D_6 + 4 D_7\right),\label{eq:alpha2}\\
\alpha_3 &= \frac{m}{32} \left( D_{14} + 16 D_{11} + 4 D_{12} + 4 D_{13} \right), \label{eq:alpha3}\\
\alpha_4&=\frac{m}{2} D_{16}, \label{eq:alpha4}\\
\alpha_5&=\frac{m}{16} \left( 8 D_{17} -D_{14} - 16 D_{11} - 4 D_{12} -4 D_{13} \right), \label{eq:alpha5}
\end{align}
the N3LO contact potential is brought in the form of Ref.~\cite{Reinert},
\begin{align}
V_{NN}^{(4)}=& D_{1}^{\prime}\left[k^{4}-4(\bm{Q} \cdot \bm{k})^{2}\right]+D_{2}^{\prime}\left[Q^{4}-\frac{1}{4}(\bm{Q} \cdot \bm{k})^{2}\right]\nonumber\\&+D_{3}^{\prime}(\bm{k} \times \bm{Q})^{2} \nonumber\\&
+\left\{D_{4}^{\prime}\left[k^{4}-4(\bm{Q} \cdot \bm{k})^{2}\right]+D_{5}^{\prime}\left[Q^{4}-\frac{1}{4}(\bm{Q} \cdot \bm{k})^{2}\right] \right. \nonumber\\& +D_{6}^{\prime}(\bm{k} \times \bm{Q})^{2} \biggr\} \bm{\sigma}_{1} \cdot \bm{\sigma}_{2}\nonumber\\&+\frac{i}{2}\left(D_{7}^{\prime} k^{2}+D_{8}^{\prime} Q^{2}\right)\left(\bm{\sigma}_{1}+\bm{\sigma}_{2}\right) \cdot(\bm{Q} \times \bm{k})\nonumber\\&
+D_{9}^{\prime}\left(-\frac{1}{4} k^{2} \bm{\sigma}_{1}\cdot \bm{k}\; \bm{\sigma}_{2} \cdot \bm{k}+4 Q^{2} \bm{\sigma}_{1} \cdot \bm{Q} \;\bm{\sigma}_{2} \cdot \bm{Q}\right)\nonumber\\&+D_{10}^{\prime} Q^{2}\left(\bm{\sigma}_{1} \cdot \bm{k}\;\bm{\sigma}_{2} \cdot \bm{k}-4 \bm{\sigma}_{1} \cdot \bm{Q}\;\bm{\sigma}_{2} \cdot \bm{Q}\right)\nonumber\\&+D_{11}^{\prime}\left(k^{2}-4 Q^{2}\right) \bm{\sigma}_{1} \cdot \bm{Q} \;\bm{\sigma}_{2} \cdot \bm{Q} \nonumber\\&+D_{12}^{\prime} \bm{\sigma}_{1} \cdot(\bm{k} \times \bm{Q})\; \bm{\sigma}_{2} \cdot(\bm{k} \times \bm{Q})\,,
\end{align}
with the following identifications
\begin{align}
D_1' &= D_1,\\
D_2' &= D_2,\\
D_3' &= D_3+D_4,\\
D_4' &= D_5,\\
D_5' &= D_6,\\
D_6' &= D_7 + D_8 + \frac{1}{16} D_{14} + D_{11} + \frac{1}{4} D_{12} + \frac{1}{4} D_{13},\\
D_7' &=D_9,\\
D_8' &=D_{10},\\
D_9' &= - 4 D_{11},\\
D_{10}' &= -\frac{1}{8} D_{14} - 2 D_{11} + \frac{1}{2} D_{12} - \frac{1}{2} D_{13},\\
D_{11}' &= - \frac{1}{8} D_{14} - 2 D_{11} - \frac{1}{2} D_{12} + \frac{1}{2} D_{13},\\
D_{12}' &= D_{15} - 2 D_{11} - \frac{1}{2} D_{12} - \frac{1}{2} D_{13} -\frac{1}{8} D_{14}.
\end{align}

\section{Induced 3N contact interactions}\label{sec:3N}
When applied to the LO 2N contact Hamiltonian, 
\begin{equation} \label{eq:vlo}
    V^{(0)}_{NN}=C_S + C_T {\bm \sigma}_1 \cdot {\bm \sigma}_2,
\end{equation}
the unitary transformation~(\ref{eq:unitaryh0}) induces additional 3N interactions \cite{Girlanda_2020} which can be viewed as a modification of the subleading 3N contact interaction entering at N4LO of the low-energy expansion \cite{Girlanda_2011},
\begin{eqnarray}\label{eq:1}
V_{3N}^{(2)}=\sum_{i j k}\bigg(&-&E_1\, \bm{k}_i^2-E_2 \,\bm{k}_i^2 \bm{\tau}_i \cdot \bm{\tau}_j\nonumber\\&-&E_3 \,\bm{k}_i^2 \bm{\sigma}_i \cdot \bm{\sigma}_j-E_4\,\bm{k}_i^2 \bm{\sigma}_i \cdot \bm{\sigma}_j \bm{\tau}_i \cdot \bm{\tau}_j\nonumber\\&-&E_5\left(3 \bm{k}_i \cdot \bm{\sigma}_i \bm{k}_i \cdot \bm{\sigma}_j-\bm{k}_i^2 \bm{\sigma}_i \cdot \bm{\sigma}_j\right)\nonumber\\&
-& E_6\left(3 \bm{k}_i \cdot \bm{\sigma}_i \bm{k}_i \cdot \bm{\sigma}_j-\bm{k}_i^2 \bm{\sigma}_i \cdot \bm{\sigma}_j\right) \bm{\tau}_i \cdot \bm{\tau}_j\nonumber\\&+&\frac{i}{2} E_7\,\bm{k}_i \times\left(\bm{Q}_i-\bm{Q}_j\right) \cdot\left(\bm{\sigma}_i+\bm{\sigma}_j\right)\nonumber\\&
+&\frac{i}{2}E_8 \,\bm{k}_i \times\left(\bm{Q}_i-\bm{Q}_j\right) \cdot\left(\bm{\sigma}_i+\bm{\sigma}_j\right) \bm{\tau}_j \cdot \bm{\tau}_k\nonumber\\&-&E_9 \,\bm{k}_i \cdot \bm{\sigma}_i \bm{k}_j \cdot \bm{\sigma}_j-E_{10}\,\bm{k}_i \cdot \bm{\sigma}_i \bm{k}_j \cdot \bm{\sigma}_j \bm{\tau}_i \cdot \bm{\tau}_j\nonumber \\
&-&E_{11}\, \bm{k}_i \cdot \bm{\sigma}_j \bm{k}_j \cdot \bm{\sigma}_i-E_{12}\,\bm{k}_i \cdot \bm{\sigma}_j \bm{k}_j \cdot \bm{\sigma}_i \bm{\tau}_i \cdot \bm{\tau}_j\nonumber\\&-& E_{13}\, \bm{k}_i \cdot \bm{\sigma}_j \bm{k}_j \cdot \bm{\sigma}_i \bm{\tau}_i \cdot \bm{\tau}_k\bigg)\nonumber\\&\equiv&\sum_{i=1}^{13}E_i\,O_i\,,
\end{eqnarray}
where $\bmk_i=\bmp'_i-\bmp_i$, $\bm{Q}_i=\frac{\bmp_i+\bmp_i'}{2}$.
Specifically, we have
\begin{equation}
    U^{\dagger} V^{(0)}_{NN} U = \sum_{i=1}^{13} \delta E_i O_i,
\end{equation}
with \footnote{We correct here some wrong factors in Ref.~\cite{Girlanda_2020}.}
{\allowdisplaybreaks
\begin{align}
\delta E_{1} &=\alpha_{1}\left(C_{S}+C_{T}\right)+\alpha_{2}\left(C_{S}-2 C_{T}\right),\label{eqs:delta1} \\
\delta E_{2} &=3\alpha_{2} C_{T}+2\alpha_{3} C_{T}-8\alpha_{4} C_{T}+2\alpha_{5} C_{T} ,\\
\delta E_{3} &=2\alpha_{1} C_{T}+\alpha_{2}\left(2 C_{S}-C_{T}\right)+\frac{2}{3}\alpha_{3}\left(2 C_{S}-C_{T}\right)\nonumber\\&+8\alpha_{4} C_{T}-2\alpha_{5} C_{T}, \\
\delta E_{4} &=\frac{2}{3}\alpha_{1} C_{T}+\frac{1}{3}\alpha_{2}\left(2 C_{S}-7 C_{T}\right)-\frac{2}{3}\alpha_{3} C_{T}\nonumber\\&+\frac{8}{3}\alpha_{4} C_{T}-\frac{2}{3}\alpha_{5} C_{T}, \\
\delta E_{5} &=2\alpha_{1} C_{T}+2\alpha_{2}\left(C_{S}-2 C_{T}\right)+\frac{2}{3}\alpha_{3}\left(2 C_{S}-C_{T}\right)\nonumber\\&+8\alpha_{4} C_{T}-2\alpha_{5} C_{T}, \\
\delta E_{6} &=\frac{2}{3}\alpha_{1} C_{T}+\frac{2}{3}\alpha_{2}\left(C_{S}-2 C_{T}\right)-\frac{2}{3}\alpha_{3} C_{T}+\frac{8}{3}\alpha_{4} C_{T}\nonumber\\&-\frac{2}{3}\alpha_{5} C_{T}, \\
\delta E_{7} &=24\alpha_{4} C_{T}, \\
\delta E_{8} &=\frac{1}{3} \delta E_{7}, \\
\delta E_{9} &=3\alpha_{1} C_{T}+3\alpha_{2}\left(C_{S}-2 C_{T}\right)+2\alpha_{3}\left(C_{S}-2 C_{T}\right)\nonumber\\&-\alpha_{4}\left(C_{S}-11 C_{T}\right)+2\alpha_{5}(C_{S}-2C_{T}), \\
\delta E_{10} &=\alpha_{1} C_{T}+\alpha_{2}\left(C_{S}-2 C_{T}\right)-\frac{1}{3}\alpha_{4}\left(3 C_{S}-15C_{T}\right), \\
\delta E_{11} &=3\alpha_{1} C_{T}+3\alpha_{2}\left(C_{S}-2 C_{T}\right)+2\alpha_{3}\left(C_{S}-2 C_{T}\right)\nonumber\\&+\alpha_{4}\left(C_{S}-11 C_{T}\right)-2\alpha_{5}(C_S-2 C_{T}), \\
\delta E_{12} &=\alpha_{1} C_{T}+\alpha_{2}\left(C_{S}-2 C_{T}\right)+\frac{1}{3}\alpha_{4}\left(3 C_{S}-15 C_{T}\right),\\
\delta E_{13}&=-16\alpha_{4} C_{T}+4\alpha_{5} C_{T}.\label{eqs:delta13}
\end{align}
}
With the specific choice for the unitary transformation encoded in Eqs.~(\ref{eq:alpha1})-(\ref{eq:alpha5}), the 3N contact LECs $E_i$ in Eq.~(\ref{eq:1}) are shifted to
\begin{equation}
    E_i \to \tilde E_i = E_i + \delta E_i,
\end{equation}
where the induced contributions $\delta E_i$ are enhanced as compared to the genuine ones $E_i$, due to the presence of the nucleon mass factor, scaling as $m\sim O(\Lambda_\chi^2/p)$ in the Weinberg counting \cite{weinberg91}, which effectively promotes them to  N3LO.
From now on, the LECs $E_i$ will be thought of as constituted only of the induced contributions, $E_i=\delta E_i$.
Thus, at N3LO the 3N contact interaction depends on five combinations of the 2N LECs $D_i$, appearing in Eqs.~(\ref{eq:alpha1})-(\ref{eq:alpha5}), which cannot be determined from 2N scattering data, but have to be fitted to experimental observables in $A>2$ systems. 

In the following we explore the sensitivity of polarization observables in low-energy $N-d$ scattering to these five combinations of LECs. Since we take the phenomenological AV18 as representative of a realistic 2N interaction, we should clarify the meaning of the LECs $C_S$ and $C_T$ in this framework. As a reasonable estimate, based on studies of universal behavior~\cite{kievskyreview}, we take them from a fit  of the LO 2N contact interaction (\ref{eq:vlo}) 
\begin{equation}
    \label{eq:vlambda}
    V^{(0)}_{NN, \Lambda}= \left[ C_S + C_T {\bm \sigma}_1 \cdot {\bm \sigma}_2 \right] Z_\Lambda (r) 
\end{equation}
to the singlets and triplets $n-p$ scattering lengths as predicted by the AV18 potential. In other words, we treat the contact potential (\ref{eq:vlambda}) as a very low-energy representation of the AV18 potential.
 In the above expression a local cutoff has been introduced, 
\begin{equation}
Z_\Lambda(r)=\int \frac{d {\bf p}}{(2 \pi)^3} {\mathrm{e}}^{i {\bf p}\cdot {\bf r}} F({\bf p}^2;\Lambda),
\end{equation}
with
\begin{equation}
  F({\bf p}^2,\Lambda)=\exp\left[-\left(\frac{{\bf p}^2}{\Lambda^2}\right)^2\right],
\end{equation}
and $\Lambda=500$~MeV.
From this procedure we get
\begin{equation}
    C_S=-66.53~{\mathrm{GeV}}^{-2}, \quad C_T=-3.47~{\mathrm{GeV}}^{-2}.
\end{equation}
The same cutoff is also used in the coordinate space expression of the induced 3N contact interaction, which becomes
\begin{eqnarray} \label{eq:induced3n}
V^{(2)}_{3N,\Lambda}&=&\sum_{i j k}  \left[  E_1 + E_2 {\bm \tau}_i \cdot {\bm \tau}_j +  \left(  E_3  +  E_4 {\bm \tau}_i \cdot {\bm \tau}_j \right)  {\bm \sigma}_i \cdot {\bm \sigma}_j\right] \nonumber \\
&& \times \left[ Z_\Lambda^{\prime\prime}(r_{ij}) + 2 \frac{Z_\Lambda^\prime(r_{ij})}{r_{ij}}\right] Z_\Lambda(r_{ik})  \nonumber \\
&& + (  E_5 + E_6 {\bm \tau}_i\cdot{\bm \tau}_j) S_{ij} \left[ Z_\Lambda^{\prime\prime}(r_{ij}) - \frac{Z_\Lambda^\prime(r_{ij})}{r_{ij}}\right] Z_\Lambda(r_{ik}) \nonumber \\
&&+ ( E_7 + E_8 {\bm \tau}_i\cdot{\bm \tau}_k) ( {\bf L}\cdot {\bm S})_{ij} \frac{Z_\Lambda^\prime(r_{ij})}{r_{ij}} Z_\Lambda(r_{ik}) \nonumber \\
&& +\left[  (E_9 +  E_{10} {\bm \tau}_j \cdot {\bm \tau}_k) {\bm \sigma}_j \cdot \hat{{\bm r}}_{ij}  {\bm \sigma}_k \cdot \hat{ {\bm r}}_{ik}   \right. \nonumber \\
&&+\left. ( E_{11} +  E_{12} {\bm \tau}_j \cdot {\bm \tau}_k+  E_{13} {\bm \tau}_i \cdot {\bm \tau}_j) {\bm \sigma}_k \cdot \hat{{\bm r}}_{ij}  {\bm \sigma}_j \cdot \hat{{\bm r}}_{ik} \right]  \nonumber \\
&& \times Z_\Lambda^\prime(r_{ij}) Z_\Lambda^\prime(r_{ik}),\label{eq:V3Ncoord}
\end{eqnarray}
where $S_{ij}$ and $ ( {\bf L}\cdot {\bm S})_{ij}$ are respectively the tensor and spin-orbit operators for particles $i$ and $j$.

\section{Low-energy p-d scattering observables within the HH method}\label{sec:methode}
In order to solve the 3-body Schr$\mathrm{\ddot{o}}$dinger equation we used the HH method, (see Refs.~\cite{hh1,hh2} for reviews).
Below the deuteron breakup threshold, the $N-d$ scattering wave function is expressed as the sum of an internal  and an asymptotic part as
\begin{equation}
\Psi_{LSJJ_z}=\Psi_C+\Psi_A\,,
\end{equation}
where the internal part $\Psi_C$ is expanded in HH as
\begin{equation}
\Psi_{C}=\sum_{\mu} c_{\mu} \Phi_{\mu}.\label{eq:psic}
\end{equation}
Here $\mu$ denotes all the quantum numbers required to fully define the basis element. The asymptotic part, $\Psi_{A}$, describes the relative motion between the nucleon and the deuteron at large distance. This latter is a linear combination of the regular and irregular solutions of the free (or Coulomb) $N-d$ Schr$\mathrm{\ddot{o}}$dinger equation, properly regularized at small distances \cite{abramowitz}.
Denoting these solutions with $\Omega_{L S J J_{z}}^{\lambda}, \lambda=R, I$ respectively, and defining
\begin{equation}
\Omega_{L S J  J_{z}}^{\pm}=i\Omega_{L S J  J_{z}}^{R} \pm \Omega_{L S J J_{z}}^{I},
\end{equation}
we have
\begin{equation} 
\Psi_{A}=\Omega_{L S J  J_{z}}^{-} +
\sum_{L^{\prime} S^{\prime}} \mathcal{S}_{L S, L^{\prime} S^{\prime}}^{J}(q) \Omega_{L^{\prime} S^{\prime} J J J_{z}}^{+}.\label{eq:PsiA}
\end{equation}
Here  $\mathcal{S}_{L S, L^{\prime} S^{\prime}}^{J}$  are the $S$-matrix elements and $q$ is defined as the modulus of the $N-d$ relative momentum. From the $S$-matrix it is possible to compute phase shift and mixing angles, from which the scattering observables are obtained. The $S$-matrix in Eq.~(\ref{eq:PsiA}) and  the coefficients $c_{\mu}$ in Eq. (\ref{eq:psic}) are obtained from the complex formulation of the  Kohn variational principle\footnote{We remind that all the states appearing in the bras $\langle \Psi |$ should be understood as $\langle \tilde \Psi|$ where $\tilde \Psi$ is the complex conjugate of $\Psi$ \cite{lucchese}.}  \cite{KIEVSKY1997125}. This principle requires that the functional
\begin{equation}
\left[\mathcal{S}_{L S, L^{\prime} S^{\prime}}^{J}(q)\right]=\mathcal{S}_{L S, L^{\prime} S^{\prime}}^{J}(q)-\frac{i}{2} \left\langle  \Psi_{L S J J_{z}}|H-E| \Psi_{L^{\prime} S^{\prime} J J_{z}}\right\rangle\label{eq:kohnvar}
\end{equation}
be stationary under variations of the trial parameters in $\Psi_{L S J J_{z}}$, with the asymptotic part normalized as
\begin{equation}
\left\langle\Omega_{L S J J_{z}}^{R}|H-E| \Omega_{L S J J_{z}}^{I}\right\rangle-\left\langle\Omega_{L S J J_{z}}^{I}|H-E| \Omega_{L S J J_{z}}^{R}\right\rangle=1 .
\end{equation}
This implies that the weights $\mathcal{S}_{L S, L S^{\prime}}^{J}$ must solve the linear system
\begin{equation}
\sum_{\tilde{L} \tilde{S}} \mathcal{S}_{L S, \tilde{L} \tilde{S}}^{J} X_{L^{\prime} S^{\prime}, \tilde{L} \tilde{S}}=Y_{L S, L^{\prime} S^{\prime}}\label{eq:systm}
\end{equation}
with
\begin{eqnarray}
X_{L S, L^{\prime} S^{\prime}}&=&\left\langle\Omega_{L S J J_{z}}^{+}|H-E| \Psi_{C}^{+} +\Omega_{L^{\prime} S^{\prime} J J_{z}}^{+}\right\rangle, \label{eq:Xls}
\\ 
Y_{L S, L^{\prime} S^{\prime}}&=&\left\langle\Omega_{L S J J_{z}}^{-}+\Psi_{C}^{-}|E- H| \Omega_{L ' S^{\prime} J J J_{z}}^{+}\right\rangle. \label{eq:Yls}
\end{eqnarray}
Here the functions $\Psi_{C}^{\pm}$ are given in Eq. (\ref{eq:psic}) with the coefficients $c_{\mu}^{\pm}$ being the solutions of

\begin{eqnarray}
&&\sum_{\mu^{\prime}}\left\langle\Phi_{\mu}|H-E| \Phi_{\mu^{\prime}}\right\rangle c_{\mu^{\prime}}^{\pm}=\nonumber\\&&-\left\langle\Phi_{\mu}|H-E| \Omega_{L S J J J_{z}}^{\pm}\right\rangle\,.\label{eq:cmu}
\end{eqnarray}
Substituting the calculated weights $\mathcal{S}_{L S, L^{\prime} S^{\prime}}^{J}$ of Eq. (\ref{eq:systm}) into Eq. ( \ref{eq:kohnvar}), it is possible to obtain a second order estimate.  In order to solve the linear system of Eq. (\ref{eq:systm}), the matrix elements of the Hamiltonian $H$ in Eqs.~(\ref{eq:Xls}), (\ref{eq:Yls}) and (\ref{eq:cmu})  have to be computed between the $\mathrm{HH}$ basis elements and the asymptotic functions.
We decompose the Hamiltonian as

\begin{equation}
H=H_{NN}+V_{3N,\Lambda}^{(0)}+V_{3N,\Lambda}^{(2)},
\end{equation}
where $H_{NN}$ is the  Hamiltonian containing the kinetic energy $T$ plus the AV18 2N interaction with Coulomb potential and $V_{3N,\Lambda}^{(0)}+V_{3N,\Lambda}^{(2)}$ contain the 3N interaction. Specifically, we consider, in addition to the induced contact interaction~(\ref{eq:induced3n}), a leading order contact interaction,
\begin{equation}
    V^{(0)}_{3N,\Lambda}= E_0 \sum_{ijk} Z_\Lambda(r_{ij}) Z_\Lambda (r_{ik}).
\end{equation}
Written $V^{(0)}_{3N,\Lambda}= E_0 V_0 $ and $V^{(2)}_{3N,\Lambda}= \sum_{i=1,13} E_i V_i$, the linear system of Eq. (\ref{eq:cmu}) results
\begin{eqnarray}
&&\sum_{\mu^{\prime}} c_{\mu^{\prime}}^{\lambda}\left\langle\Phi_{\mu}\left|H_{NN}+\sum_{i=0,13} E_{i} V_{i}-E\right| \Phi_{\mu^{\prime}}\right\rangle=\nonumber\\&&-\left\langle\Phi_{\mu}\left|H_{NN}+\sum_{i=0,13} E_{i} V_{i}-E\right| \Omega_{L S J J J_{z}}^{\lambda}\right\rangle,
\end{eqnarray}
which can be put in the matrix form
\begin{eqnarray}
&&\sum_{\mu^{\prime}}\left[\left(H_{NN}\right)_{\mu \mu^{\prime}}+\sum_{i=0,13} E_{i}\left(V_{i}\right)_{\mu \mu^{\prime}}-E N_{\mu \mu^{\prime}}\right] c_{\mu^{\prime}}^{\lambda}=\nonumber\\&&-\left(H_{NN}\right)_{\mu \lambda}+\sum_{i=0,13} E_{i}\left(V_{i}\right)_{\mu \lambda}-E N_{\mu \lambda}.\label{eq:Hmumu'}
\end{eqnarray}
Here  $\left(\right)_{\mu \mu^{\prime}}$ denotes the matrix elements of the considered operator between the corresponding basis states. 
The contact potential matrix can be computed as a linear combination of several matrices, one for $V^{(0)}$ and one for each operator appearing in $V^{(2)}$. With the proper corresponding  LECs, these matrices can be constructed once and used for all purposes. Their size is approximately 2000$\times$2000, making the computation feasible in few seconds for each channel on an ordinary desktop. With these dimensions the observables are calculated well inside a 1\% accuracy \cite{nogga,kievsky2001,kievsky1998}. 
A specific set of LECs can be utilized to compute the associated  $S$-matrix for each $J^{\pi}$ state using the Kohn variational principle, from which the observables at a specific energy $E$ can be obtained.
In order to do this, we compute the $N-d$ transition matrix $M$, which is composed of the Coulomb amplitude $ f_{c}(\theta_{\mathrm{cm}})$  and a nuclear term, $\theta_{\mathrm{cm}}$ being the center-of-mass scattering angle, as

\begin{align} \label{eq:Mmatrix}
M_{\nu \nu^{\prime}}^{S S^{\prime}}(\theta_{\mathrm{cm}})=&  f_{c}(\theta_{\mathrm{cm}}) \delta_{S S^{\prime}} \delta_{\nu \nu^{\prime}}+\frac{\sqrt{4 \pi}}{q} \sum_{L L^{\prime}J} \sqrt{2 L+1}\nonumber\\
&(L 0 ,\,S \nu \mid J \nu)\left(L^{\prime} M^{\prime},\,S^{\prime} \nu^{\prime} \mid J \nu\right) \nonumber\\
& {\mathrm{e}}^{i \left(\sigma_{L}+\sigma_{L^{\prime}}-2 \sigma_{0}\right)} T_{L S, L^{\prime} S^{\prime}}^{J} Y_{L^{\prime} M^{\prime}}(\theta_{\mathrm{cm}}, 0).
\end{align}
Here the matrix $M_{\nu \nu^{\prime}}^{S S^{\prime}}(\theta_{\mathrm{cm}})$ is a $6 \times 6$ matrix corresponding to the couplings of the spin 1 of the deuteron and the spin $1 / 2$ of the third nucleon, to $S, S^{\prime}=1 / 2$ or $3 / 2$ with projections $\nu$, $\nu^{\prime}$. The quantum numbers $L, L^{\prime}$ are the relative orbital angular momentum between the deuteron and the third particle and $J$ is the total angular momentum.
The matrix elements $T_{L S, L^{\prime} S^{\prime}}^{J}$ form the $T$-matrix of a Hamiltonian containing the nuclear plus
Coulomb interactions. Note that the $T$-matrix can be related with the $S$-matrix of Eq.~(\ref{eq:PsiA}) by $S=1-2i\pi T$.
Finally,  $\sigma_L$ are the Coulomb phase–shifts. The effect of other components of the electromagnetic interaction are discussed in Ref.~\cite{kievsky_em}.

\section{Fit results}\label{sec:results}
The  observables used in the fitting procedure are the $p-d$ differential cross section, the two vector analyzing powers $A_{y}$ and $i T_{11}$, the three tensor analyzing powers $T_{20}, T_{21}$, $T_{22}$ and the doublet and quartet  $n-d$ scattering lengths. In particular we determine the leading contact LEC $E_{0}$ from the experimental triton binding energy. Then, we  fit the experimental doublet and quartet $n-d$ scattering lengths \cite{scatteringlenght1,scatteringlenght2} and the six $p-d$ scattering observables at center-of-mass energy $E_{\mathrm{cm}}=2\, \mathrm{MeV}$ \cite{shimizu}, amounting to $282$ experimental data. The theoretical observables are calculated solving Eqs. (\ref{eq:systm}) and (\ref{eq:cmu}), then the obtained $S$-matrix  is used to calculate the transition matrix $M$ of Eq.~(\ref{eq:Mmatrix}), from which the observables are directly calculated \cite{GLOCKLE1996107}. At the energy considered, states up to $L=2$ are calculated using the full Hamiltonian, whereas for $L>2$  the three-body potential was neglected due to its short-range character (see also Ref.~\cite{tornow}), while the strong two-body potential was included up to a maximum value of $L=6$ in the partial wave expansion of the observables, which is enough at the energy of interest.

For the differential cross section we include in the $\chi^{2}$  definition an overall normalization factor $Z$ of the data points, i.e

\begin{equation}
\chi^{2}=\sum_{i} \frac{\left(d_{i}^{\exp } / Z-d_{i}^{\mathrm{th}}\right)^{2}}{\left(\sigma_{i}^{\exp } / Z\right)^{2}},\label{eq:chi2}
\end{equation}
with $Z$ obtained from the minimization condition as
\begin{equation}
Z=\frac{\sum_{i} d_{i}^{\exp } d_{i}^{\mathrm{th}} /\left(\sigma_{i}^{\exp }\right)^{2}}{\sum_{i}\left(d_{i}^{\mathrm{th}}\right)^{2} /\left(\sigma_{i}^{\exp }\right)^{2}}.\label{eq:Z}
\end{equation}
In Eqs. (\ref{eq:chi2}) and (\ref{eq:Z}) $d_{i}^{\exp/\mathrm{th}}$ are the experimental data points and their theoretical predictions, while $\sigma_{i}^{\exp }$ is the experimental error. In our study we have checked that $Z$ never differs from 1 by more than $2 \%$ \cite{PhysRevC.64.024002}. For the other observables, we treat the normalization $Z=1.00 \pm 0.01$ as an additional experimental datum since, according to Ref. \cite{shimizu}, the systematic uncertainty is estimated as $1 \%$.

For an initial random set of the five $\alpha_i$ parameters of  Eqs.~(\ref{eq:alpha1})-(\ref{eq:alpha5}), we solve the scattering problem and calculate the corresponding observables. Using the POUNDerS algorithm \cite{pounders} we start an iterative procedure to minimize the global $\chi^{2} /$ d.o.f. of the data set description. Using different initial random input of $\alpha_i$  values, we repeat the algorithm trying to localize the deepest mimimum. This amounts to $\chi^2$/d.o.f.\ =1.7, of the same quality as the most accurate multiparameter fits to the same data performed so far \cite{contact19}. 
\begin{figure*}
\begin{center}
\includegraphics[scale=0.6]{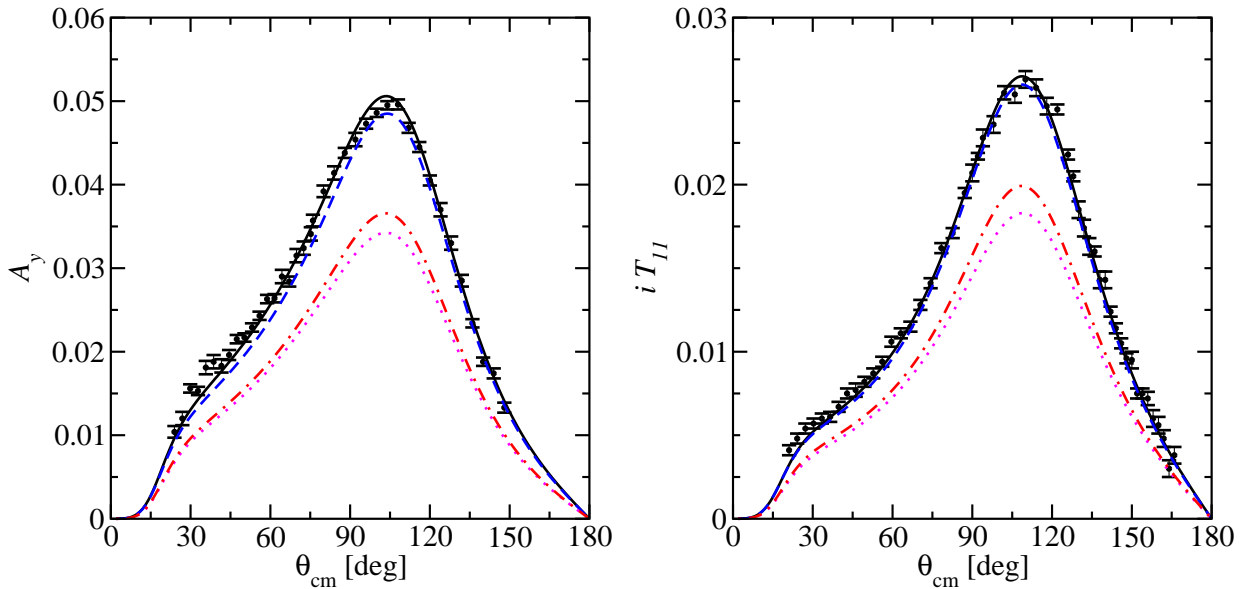}
\caption{Proton and deuteron analyzing power in $\vec{p}-d$ and $\vec{d}-p$ scattering at $E_{\mathrm{cm}}=2$~MeV. The full (black) lines result from a global 5-parameter fit, the dashed (blue) lines from a 3-parameter fit excluding the ${\bf P}$-dependent 2N interaction, the dotted (pink) lines are the predictions from the 2N AV18 potential, while the dashed-dotted (red) lines are the predictions including also the 3N Urbana IX interaction. Experimental data are from Ref.~\cite{shimizu}.}\label{fig:ay}
\end{center}
\end{figure*}

Fig.~\ref{fig:ay} shows the best fit curve for the $A_y$  and $i T_{11}$ analyzing power in $\vec{p}-d$ and $\vec{d}-p$ scattering, compared to the predictions from the purely 2N AV18 interaction and from the addition of the  Urbana IX  3N interaction. We conclude that the effective N3LO induced 3N contact interaction allows to solve the long-standing $A_y$ problem.  Also the description of the vector analyzing power $i T_{11}$ is drastically improved.  

We also show in the same figure the best fit curve obtained from a 3-parameter fit which does not include the $\alpha$-parameters of the ${\bf P}$-dependent N3LO 2N contact interaction, i.e. with $\alpha_4=\alpha_5=0$, in order to assess the relevance of the LECs $D_{16}$ and $D_{17}$, which were never considered before. No spin-orbit operators, of the kind proposed in Ref.~\cite{Kievsky_1999}, are present in this latter case, and the minimum  $\chi^2$/d.o.f.\ increases to 2.3.
\begin{figure*}
\begin{center}
\includegraphics[scale=0.7]{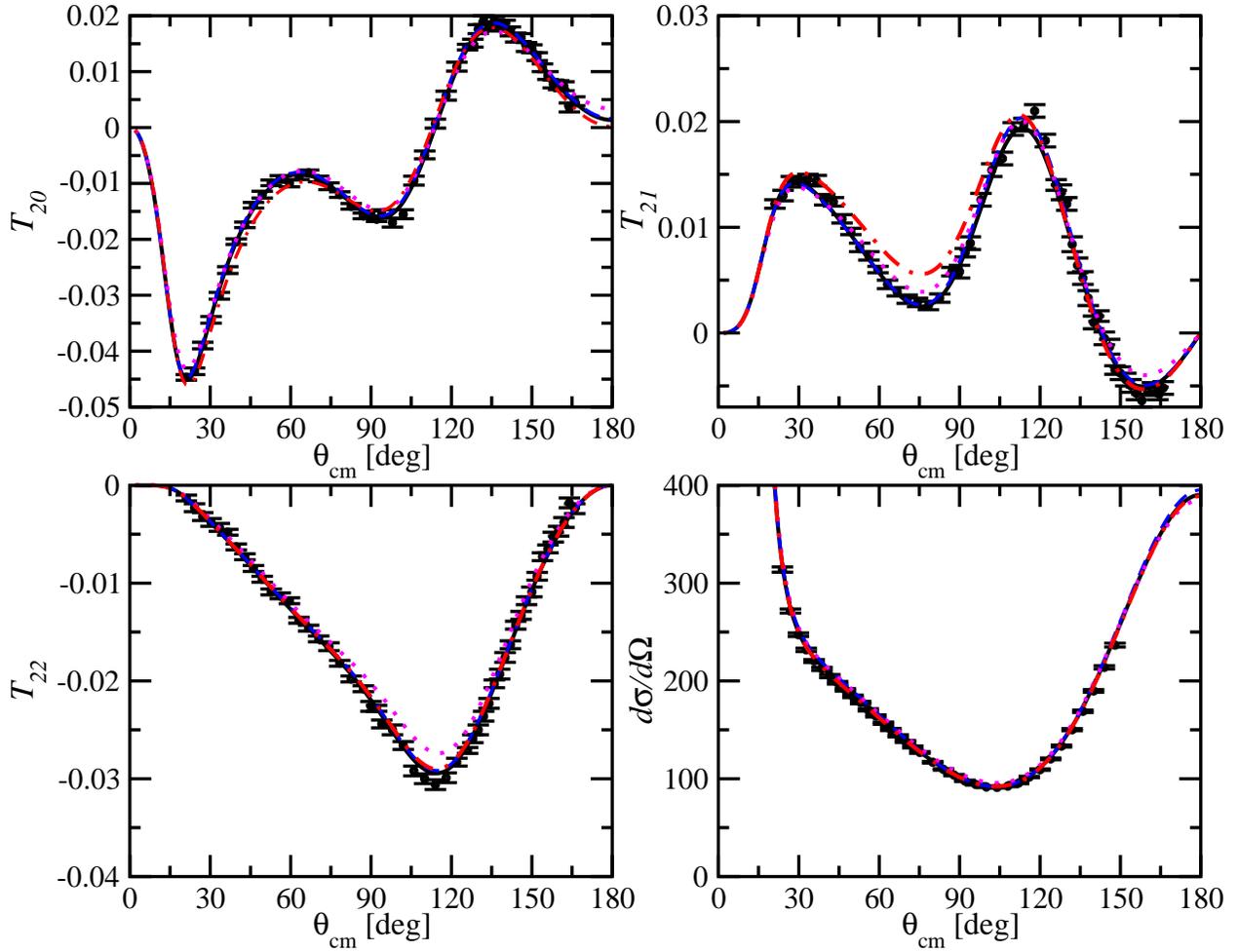}
\caption{Same as Fig.~\ref{fig:ay} but for $T_{20}$, $T_{21}$, $T_{22}$ tensor observables in $\vec{d}-p$ scattering and for the unpolarized differential cross-section  at $E_{\mathrm{cm}}=2$~MeV. }\label{fig:tdeut}
\end{center}
\end{figure*}

In Fig.~\ref{fig:tdeut} we show the same curves for the tensor analyzing powers of $\vec{d}-p$ elastic scattering and for the differential cross-section. By inspection of the figures, we can conclude that all the observables are nicely reproduced.

\begin{table}
\centering
 \begin{tabular}{|c|c|c|}
\hline
Fitting procedure &  5-param. & 3-param.\\
\hline
$\chi^2$/d.o.f. & 1.7  & 2.3\\
\hline
$e_0$ & 0.685 & -1.570\\
$\tilde \alpha_1 C_S$ & 1.410  &-3.611 \\
$\tilde \alpha_2 C_S$ & 0.211  & -0.483\\ 
$\tilde \alpha_3 C_S$ & -0.370 & 0.209\\
$\tilde \alpha_4 C_S$ & 1.735  & 0\\
$\tilde \alpha_5 C_S$ & 2.266  & 0\\
\hline
$^2 a_{nd}$ [fm] & 6.31  & 6.32 \\
$^4 a_{nd}$  [fm] & 0.648  & 0.647\\
\hline
 \end{tabular}
 \caption{Results of the 5-parameters and 3-parameters fits, the latter one obtained ignoring the ${\bf P}$-dependent 2N contact interaction, i.e. setting $\alpha_4=\alpha_5=0$. See text for more explanations.} \label{tab:fits}
 \end{table}
The fitted parameters $\alpha_i$ are displayed in Table~\ref{tab:fits}, together with the corresponding values of the LO 3N contact LEC $E_0$ in units as dictated by naive dimensional analysis \cite{nda1,nda2}, i.e. 
\begin{equation}
e_0 = E_0 F_\pi^4 \Lambda, \quad \tilde \alpha_i = \alpha_i F_\pi^4 \Lambda^3,
\end{equation}
where $F_\pi=92.4$ MeV is the pion decay constant.
Also shown in the table are the doublet and quartet $n-d$ scattering lengths, to be compared with the experimental values $^2a_{nd}=(0.645\pm 0.003 \pm 0.007)$~fm \cite{scatteringlenght1} and $^4a_{nd} = (6.35 \pm 0.02)$~fm \cite{scatteringlenght2}.
It is interesting to observe that the fitted 3N interaction parameters are of a natural size for a N3LO contribution. In order to see this, we can translate the values of the $\alpha_i$'s into combinations of the N3LO 2N LECs $D_i$'s using  Eqs.~(\ref{eq:alpha1})-(\ref{eq:alpha5}).  This is done in Table \ref{tab:Destimate} for the two fitting procedures. As a reference, we report in the same table the corresponding combinations of LECs obtained from 2N data in Ref.~\cite{Machleidt_2011}, and used in the Idaho N3LO 2N chiral potential with $\Lambda=500$~MeV. The comparison of the actual values has little meaning, also due to the hybrid character of our calculation. However it is interesting to observe that  the orders of magnitude are the same. In particular, for the 5-parameter fit, the LECs combinations are not larger than those obtained in the Idaho N3LO chiral potential.
 \begin{table}
\centering
 \begin{tabular}{|c|c|c|c|}
 \hline
 & 5-param. & 3-param. &  Ref.\cite{Machleidt_2011} \\
 \hline
$D_{16}$ & -0.610 & 0& -\\
$D_{17}$ &-0.536 & -0.181 & -\\ 
$16D_1+D_2+4D_3$ & -3.96 & 10.86& 6.41\\
$16D_5+D_6+4D_7$ & -0.593&  1.21 &4.05\\
$D_{14}+16D_{11}+4D_{12}+4D_{13}$ & 2.08 & -1.44 & -3.04\\
\hline
 \end{tabular}
 \caption{Estimation of some N3LO LECs combinations, the $D_i$ are in units of $10^4$ GeV$^{-4}$. In the second column we show the values obtained from the  5-parameter fit, in the third  one the estimates obtained from the 3-parameter fit, while the last column shows the values obtained in Ref. \cite{Machleidt_2011} and used for the Idaho N3LO 2N potential with $\Lambda=500$~MeV. } \label{tab:Destimate}
 \end{table}
 We advocate that, were those combinations fitted in the $A=3$ system, the $A_y$ puzzle would be solved at N3LO. However this remains to be seen explicitly in a consistent chiral calculation.
\section{Conclusions}\label{sec:conclusions}
A suitable choice of unitary transformation allows to reduce the number of LECs parametrizing the N3LO 2N contact interaction to twelve. This procedure generates a 3N interaction depending on five unconstrained LECs. In the present paper we examined the effect of this induced 3N interaction on polarization observables of $p-d$ scattering below the breakup threshold. We showed that the LECs can be adjusted allowing to solve the long-standing $A_y$ puzzle.

The induced 3N interaction can be thought of as a specific off-shell extension of the 2N interaction, leaving the 2N observables unchanged. Such off-shell extension of the 2N potentials  were considered in the past (see e.g. Ref.~\cite{Doleschall2000} ) and found to have a prominent role in the $N-d$ $A_y$ puzzle \cite{Doleschall2004}. We remark in passing that a satisfactory fit (with $\chi^2$/d.o.f.~=1.8) can be obtained even without including any 3N interaction except for the induced one, i.e. with $E_0=0$.
We emphasize that the novelty of our proposal lies in the identification  of its precise form in the context of a systematic low-energy expansion, where it starts to contribute at N3LO.
 This statement has also a quantitative content, despite all the limitations of our hybrid calculation, in light of the comparison of the magnitudes of the involved LECs with those inferred within the ChEFT  framework of the 2N interaction, as shown in Table~\ref{tab:Destimate}.
 
 Of course it will be interesting to repeat the above analysis in a fully consistent ChEFT framework for  2N and 3N interactions. In this respect, also the induced 3N interaction from the unitary transformation of the one-pion exchange 2N potential has to be taken into account. To the best of our knowledge such contribution, first worked out in Ref.~\cite{Girlanda_2020}, has never been considered in the literature so far. In the present work it was implicitly taken into account through the values of the LECs $C_S$ and $C_T$,  by considering a pionless representation of the AV18 potential.
 It will be also necessary to explore the energy dependence of the predicted $p-d$ scattering observables and confront it with experimental data. Such exploration has been pursued  in Ref.~\cite{contact19} to  energies lower than $E_{\mathrm{cm}}=2$~MeV using a restricted form for the subleading 3N contact interaction, leading to quite satisfactory results. 
Finally, the same shuffling of contact operators between the 2N and 3N sectors applies to the pionless formulation of the EFT. The counting of the induced 3N operators examined in the present paper should follow from the corresponding counting of the 2N operators. A further peculiarity in this case is the promotion of the 3N force to LO. Thus the appropriate counting should be re-examined in this perspective (see also Ref.~\cite{implications}).
Work along the lines outlined above is deferred to forthcoming investigations.

\end{document}